\documentstyle[12pt]{article}
\begin{document}
\baselineskip=24pt
\def\rd{{\rm d}}
\newcommand{\Lamb}{\Lambda_{b}}
\newcommand{\Lamc}{\Lambda_{c}}
\newcommand{\dsp}{\displaystyle}
\newcommand{\nn}{\nonumber}
\newcommand{\dfr}[2]{ \displaystyle\frac{#1}{#2} }
\newcommand{\Lag}{\Lambda \scriptscriptstyle _{ \rm GR} } 
\newcommand{\pa}{p\parallel}
\newcommand{\pe}{p\perp}
\newcommand{\pet}{p\top}
\newcommand{\paa}{p'\parallel}
\newcommand{\pee}{p'\perp} 
\newcommand{\pete}{p'\top} 
\renewcommand{\baselinestretch}{1.5}
\begin{titlepage}
\vspace{-20ex}
\vspace{1cm}

\centerline{\Large\sf $\Lambda_b \rightarrow \Lambda_c P(V)$ Nonleptonic Weak Decays}
\vspace{6.4ex}
\centerline{\large\sf  	X.-H. Guo$^{1,2}$}
\vspace{3.5ex}
\centerline{\sf 1. Department of Physics and Mathematical Physics,}
\centerline{\sf and Special Research Center for the Subatomic Structure of
Matter,}
\centerline{\sf University of Adelaide, SA 5005, Australia}
\centerline{\sf 2. Institute of High Energy Physics, Academia Sinica,
Beijing, China}
\vspace{6ex}
\begin{center}
\begin{minipage}{5in}
\centerline{\large\sf 	Abstract}
\vspace{1.5ex}
\small {The two-body nonleptonic weak decays of $\Lambda_b \rightarrow 
\Lambda_c P(V)$
($P$ and $V$ represent pseudoscalar and vector mesons respectively)
are analyzed in two models, one is the Bethe-Salpeter (B-S) model and the other
is the hadronic wave function model. The calculations are carried out
in the factorization approach. The obtained results are compared with
other model calculations.}

\end{minipage}
\end{center}

\vspace{1cm}

\end{titlepage}
\vspace{0.2in}
{\large\bf I. Introduction}
\vspace{0.2in}

Recently, experimental measurements for the heavy baryon $\Lambda_b$ begin to
be available. For example, OPAL has measured some physical
quantities for $\Lambda_b$ such as its lifetime and the product 
branching ratio for the inclusive semileptonic decay 
$\Lambda_b \rightarrow \Lambda l^- \bar{\nu} X$ \cite{opal}. Furthermore, 
the measurements for the nonleptonic decay of $\Lambda_b$ also appeared.
This is the well-known process $\Lambda_b \rightarrow \Lambda J/\psi$. 
The discrepancy between the measurements made by UA1 \cite{ua1} and CDF, LEP
\cite{cdf2}\cite{lep} has been settled down by the new measurement from
CDF \cite{cdf1}. However, comparing with the data of D, B and $\Lambda_c$ the
data for $\Lambda_b$ is still very limited. But we certainly expect more and
more data coming out in the near future. 

On the other hand, there have been also some progress in the theoretical
study on heavy baryon decays. In comparison with the case of heavy mesons
the situation for heavy baryons becomes more complicated since there are
three quarks in a baryon instead of two in a meson. Fortunately the
establishment of the heavy quark effective theory (HQET) \cite{wise}
makes the study on heavy flavor hadrons simplier since the 
HQET can reduce the independent number of weak transition form factors. 
It can be shown that  in the leading order of $1/m_Q$ expansion 
there is only one 
form factor, the Isgur-Wise function, for $\Lambda_b \rightarrow \Lambda_c$ 
weak transition.
Furthermore, HQET can also be applied to relate some nonleptonic decay 
processes in the heavy quark limit. For instance, the decay widths of 
$\Lambda_b \rightarrow \Lambda_c D_s$ and $\Lambda_b \rightarrow \Lambda_c
D_{s}^{*}$ are related to each other by the 
heavy quark symmetry \cite{grinstein}.
The decay widths of these two processes are expressed by two common  
scalar functions in the heavy quark limit.

Although the 
heavy quark symmetry can be used to simplify the physical processes
where heavy hadrons are involved, in most cases HQET itself cannot give
the final phenomenological predictions for the decay properties.
Hence one still has to adopt nonperturbative QCD
models in the end. Among them we have, for instance, QCD sum rules, the
Bethe-Salpeter (B-S) equation, chiral perturbation theory, potential model,
bag model, instanton model, relativistic and nonrelativistic quark model, etc.
By applying these models one can calculate the weak transition form factors
such as $\Lambda_b \rightarrow \Lambda_c$. Consequently the semileptonic decay
widths are drawn out directly since the lepton pair can be extracted from
the hadronic weak transition form factor.

For nonleptonic decays, things become much more complicated. To simplify
the calculations the factorization assumption is applied so that one of the
currents in the nonleptonic decay Hamiltonian is factorized out and
generates a meson. Thus the decay amplitude of the two body nonleptonic decay
becomes the product of two matrix elements, one is related to the decay
constant of the factorized meson and the other is the weak transition
matrix element between two hadrons. There have been some discussions about the
plausibility of the factorization approach. In the energetic weak decays the 
quark pair generated by one current in the weak 
Hamiltonian moves very fast away
from the weak interaction point. Therefore, by the time this quark pair
hadronizes into a meson it is far away from other quarks and it almost
does not interact with the remaining quarks. Hence this quark pair is
factorized out and generates a meson. This argument is based on the ideas
of ``color transparency'' given by Bjorken \cite{bjorken}. Dugan and
Grinstein proposed a formal proof for factorization approach by constructing
a large energy effective theory \cite {dugan}. 
It is shown that when the energy of the
generated meson is very large the meson can be factorized out and the deviation
from the factorization amplitude is suppressed by the energy of the factorized
meson. In the $\Lambda_b \rightarrow \Lambda_c P(V)$ decays
the W-exchange diagram is also involved
besides factorization diagrams. However,
it is argued that in the bottom baryon case the W-exchange diagram's
contribution is suppressed by one order caused by 
a factor $32\pi |\psi (0)|^2/m_{b}^{3}$
with respect to the spectator diagram \cite{cheng0}\cite{cheng1}. 
From the above arguments the factorization
approach is a good approximation for $\Lambda_b$ nonleptonic decays. 
In fact in the B meson nonleptonic
decays it has been shown that the factorization
approach works well since it leads to theoretical predictions which in general
are consistent with experimental data \cite{stech}.

In the
nonrelativistic quark model Cheng calculated the decay widths for many
processes of $\Lambda_b$ in the factorization approach \cite{cheng1}. 
Actually the factorization
contribution had been considered by Mannel and Roberts \cite{mannel}. By
simply applying the Isgur-Wise function for $B \rightarrow D$ to  
$\Lambda_b \rightarrow \Lambda_c$ they gave the Cabibbo-allowed decay branching
ratios. Because the light degrees of 
freedom in a heavy meson and a heavy baryon has
different dynamics, their Isgur-Wise functions should also be different. 
For instance, in the B-S equation model it is shown that the Isgur-Wise
function for $\Lamb \rightarrow \Lamc$ drops faster that that for 
$B \rightarrow D$ \cite{bs}. 
In the factorization approach, the essential point is the weak
transition form factors for $\Lambda_b \rightarrow \Lambda_c$. In our previous
papers we dealt with this transition in the B-S equation approach to the
leading order in $1/m_Q$ (Q=b or c) expansion \cite{bs} and in the hadronic 
wave function
model in the infinite momentum frame (IMF) to the order $1/m_Q$ \cite{gk}. 
It is the motivation of the present paper to apply these results to the
nonleptonic decays of $\Lambda_b$.

When the quark mass is very heavy comparing with the QCD scale
$\Lambda_{QCD}$, the light degrees of freedom in a heavy baryon  
$\Lambda_{Q}$  becomes blind to the flavor and spin quantum
numbers of the heavy quark because of the $SU(2)_f \times 
SU(2)_s$ symmetries. Therefore, the angular momentum  
and flavor quantum numbers 
of the light degrees  of freedom (the light diquark) become good quantum 
numbers which can be used to classify heavy baryons. It is thus reasonable
to assume that the heavy baryon $\Lambda_{Q}$ is composed of a heavy quark
and a scalar diquark. In this picture, the
three body system is simplified to two body system. 
Based on this simple picture, we established the B-S equation for
$\Lambda_{Q}$ in the heavy quark limit and solved out the Isgur-Wise
function \cite{bs}. Also in this two body-system picture, in IMF we 
combined the Drell-Yan type overlap integrals with the results from the HQET 
to calculate all the six form factors which describe the transition
$\Lambda_b \rightarrow \Lambda_c$ to the order of $1/m_Q$.

The remainder of this paper is organized as follows. In Sect. II we
give the formulation to deal with the $\Lamb \rightarrow \Lamc P(V)$
decays. Then in Sect. III we present the results from the B-S approach and
the hadronic wave function model. Other model calculations are also listed
for comparison. Finally, Sect. VI is served for summary and discussions.

\vspace{0.2in}
{\large\bf II. Formulation}
\vspace{0.2in}

In this section we briefly review the standard formulation for calculating
the decays $\Lamb \rightarrow \Lamc P(V)$.

The Hamiltonian describing the two body nonleptonic decays of a bottom
baryon reads
\begin{equation}
H_{eff}=\frac{G_F}{\sqrt{2}}V_{cb}V^{*}_{UD}(a_1 O_1 +a_2 O_2),
\label{1}
\vspace{2mm}
\end{equation}
with $O_1=(\bar D U)(\bar c b)$ and $O_2=(\bar c U)(\bar D b)$, where $U$
and $D$ are the fields for light quarks involved in the decay, and 
$(\bar q_1 q_2)=\bar q_1 \gamma_\mu (1-\gamma_5) q_2$ is understood.
The parameters $a_1$ and $a_2$ are treated as free parameters since they
involve hadronization effects. In literatures usually $a_1$ and $a_2$
are expressed in terms of the QCD coefficients $c_1$ and $c_2$
\begin{eqnarray}
a_1&=&c_1+\zeta c_2, \nn \\
a_2&=&c_2+\zeta c_1,
\label{2}
\vspace{2mm}
\end{eqnarray}
where $c_1 (m_b)=1.11$ and $c_2 (m_b)=-0.26$. In the naive factorization
approach $\zeta=1/N_c$ with $N_c=3$. However, because of the color-octet
contributions the value of $\zeta$ differs from $1/3$. In the charm decays
$\zeta \sim 0$ and in the bottom case it is still not very clear. In ref. 
\cite{cheng2} it is proposed that in the bottom case $\zeta \sim 1/2$.
The values of $a_1$ and $a_2$ need to be clarified when more data on
the bottom  hadrons are available. In the present work we simply treat them
as free parameters.

The general form for the amplitudes of $\Lambda_b \rightarrow \Lambda_c P(V)$
are \cite{cheng1}
\begin{eqnarray}
M(\Lambda_b \rightarrow \Lambda_c P)&=&i\bar{u}_{\Lambda_c}(p_{\Lambda_c})
(A+B\gamma_5)
u_{\Lambda_b}(p_{\Lambda_b}), \nn \\
M(\Lambda_b \rightarrow \Lambda_c V)&=&\bar{u}_{\Lambda_c}(p_{\Lambda_c})
\epsilon^{*\mu}
[A_1\gamma_\mu \gamma_5+A_2 (p_{\Lambda_c})_{\mu}\gamma_5+B_1\gamma_\mu
+B_2 (p_{\Lambda_c})_\mu]u_{\Lambda_b}(p_{\Lambda_b}), \nn\\
& &
\label{3}
\vspace{2mm}
\end{eqnarray}
where $u_{\Lambda_c}$, $u_{\Lambda_b}$ are dirac spinors of 
$\Lambda_c$, $\Lambda_b$ 
respectively and $\epsilon_\mu$ is the polarization vector of the emitted
vector meson.

In the factorization approach the amplitudes for $\Lambda_b \rightarrow 
\Lambda_c P(V)$
is 
\begin{equation}
M^{fac}(\Lambda_b \rightarrow \Lambda_c P(V))=\frac{G_F}{\sqrt{2}}
V_{cb}V^{*}_{UD}
a_1 <P(V)|A_\mu(V_\mu)|0> <\Lambda_c (p_{\Lambda_c)}|J^\mu|\Lambda_b 
(p_{\Lambda_b})>,
\label{4}
\vspace{2mm}
\end{equation}
where $J_\mu$ is the $V-A$ weak current and $<0|A_\mu(V_\mu)|P(V)>$ are related
to the decay constants of the pseudoscalar meson  or vector meson  by
\begin{eqnarray}
<0|A_\mu|P>&=&if_P q_\mu, \nn\\
<0|V_\mu|V>&=&f_V m_V \epsilon_{\mu},
\label{5}
\vspace{2mm}
\end{eqnarray}
where $q_\mu$ is the momentum of the emitted meson from W-boson and the
normalization is chosen so that $f_\pi=132$MeV. 
It is noted that in the two-body nonleptonic weak decays
$\Lambda_b \rightarrow \Lambda_c P(V)$ there is no contribution from the
$a_2$ term since such a term corresponds to the transition of $\Lamb$ to
a light baryon instead of $\Lamc$.

The matrix element for $\Lambda_b \rightarrow \Lambda_c$ can be expressed
as the following on the ground of Laurance invariance
\begin{eqnarray}
<\Lambda_c (p_{\Lambda_c)}|J_\mu|\Lambda_b (p_{\Lambda_b})>&=&
\bar{u}_{\Lambda_c}(p_{\Lambda_c})[f_1(q^2)\gamma_\mu+if_2(q^2)\sigma_{\mu\nu}
q^\nu+f_3(q^2)q_\mu \nn \\
& &-(g_1(q^2)\gamma_\mu+ig_2(q^2)\sigma_{\mu\nu}
q^\nu+g_3(q^2)q_\mu)\gamma_5]
u_{\Lambda_b}(p_{\Lambda_b}), 
\label{6}
\vspace{2mm}
\end{eqnarray} 
where $f_i$, $g_i$ (i=1,2,3) are the Laurance scalars. Alternatively, in the
heavy baryon case we the above matrix element can be expressed in terms of the
velocities of $\Lamb$ and $\Lamc$,
\begin{eqnarray}
<\Lambda_c (v_{\Lambda_c)}|J_\mu|\Lambda_b (v_{\Lambda_b})>&=&
\bar{u}_{\Lambda_c}(v_{\Lambda_c})[F_1(\omega)\gamma_\mu+F_2(\omega)
v_{\Lambda_b \mu}+F_3(\omega)v_{\Lambda_c \mu} \nn \\
& &-(G_1(\omega)\gamma_\mu+G_2(\omega)
v_{\Lambda_b \mu}+G_3(\omega)v_{\Lambda_b \mu})\gamma_5], 
\label{7}
\vspace{2mm}
\end{eqnarray}
where $\omega=v_{\Lamc}\cdot v_{\Lamb}$. The relations between $f_i, \; g_i$
and $F_i, \; G_i$ are 
\begin{eqnarray}
f_1&=&F_1+\frac{1}{2}(m_{\Lamb}+m_{\Lamc})\left( \frac{F_2}{m_{\Lamb}}
+\frac{F_3}{m_{\Lamc}} \right), \nn\\
f_2&=&\frac{1}{2}\left( \frac{F_2}{m_{\Lamb}}
+\frac{F_3}{m_{\Lamc}} \right), \nn\\
f_3&=&\frac{1}{2}\left( \frac{F_2}{m_{\Lamb}}
-\frac{F_3}{m_{\Lamc}} \right), \nn\\
g_1&=&G_1-\frac{1}{2}(m_{\Lamb}-m_{\Lamc})\left( \frac{G_2}{m_{\Lamb}}
+\frac{G_3}{m_{\Lamc}} \right), \nn\\
g_2&=&\frac{1}{2}\left( \frac{G_2}{m_{\Lamb}}
+\frac{G_3}{m_{\Lamc}} \right), \nn\\
g_3&=&\frac{1}{2}\left( \frac{G_2}{m_{\Lamb}}
-\frac{G_3}{m_{\Lamc}} \right). 
\label{8}
\vspace{2mm}
\end{eqnarray}
In the heavy quark limit $m_Q \rightarrow \infty$  
$$F_1=G_1=\xi(\omega),\;\; \;\; F_2=F_3=G_2=G_3=0,$$
where $\xi(\omega)$ is the Isgur-Wise function. 

The decay widths and the up-down asymmetries
for $\Lamb \rightarrow \Lamc P(V)$ are available in literatures
\cite{cheng1}\cite{tuan}. 
\begin{eqnarray}
\Gamma(\Lamb \rightarrow \Lamc P)&=&\frac{p_c}{8\pi}\left[ \frac{(m_{\Lamb}
+m_{\Lamc})^2-m_{P}^{2}}{m_{\Lamb}^2}|A|^2+\frac{(m_{\Lamb}
-m_{\Lamc})^2-m_{P}^{2}}{m_{\Lamb}^2}|B|^2 \right], \nn\\
\alpha(\Lamb \rightarrow \Lamc P)&=&-\frac{2\delta Re(A^*B)}
{|A|^2+\delta^2|B|^2},
\label{kin1}
\vspace{2mm}
\end{eqnarray}
where $p_c$ is the c.m. momentum and $\delta=\sqrt{(E_{\Lamc}-m_{\Lamc})/
(E_{\Lamc}+m_{\Lamc})}$. A and B are related to the form factors by
\begin{eqnarray}
A&=&\frac{G_F}{\sqrt{2}}V_{cb}V^{*}_{UD}a_1 f_P (m_{\Lamb}-m_{\Lamc}) 
f_1(m_{P}^{2}), \nn\\
B&=&\frac{G_F}{\sqrt{2}}V_{cb}V^{*}_{UD}a_1 f_P (m_{\Lamb}+m_{\Lamc}) 
g_1(m_{P}^{2}).
\label{kin2}
\vspace{2mm}
\end{eqnarray}
\begin{eqnarray}
\Gamma(\Lamb \rightarrow \Lamc V)&=&\frac{p_c}{8\pi}
 \frac{E_{\Lamc} + m_{\Lamc}}{m_{\Lambda_{b}}}
\left[2(|{S}|^{2} + |{P_{2}}|^{2}) + \frac{E_{V}^{2}}
 {m_{V}^{2}}(|{S + D}|^{2} + |{P_{1}}|^{2}) \right], \nn\\
\alpha(\Lamb \rightarrow \Lamc V)&=&\frac{4m_{V}^{2}Re(S^{*}P_{2})
+2E_{V}^{2}Re(S+D)^{*}P_{1}}{2m_{V}^{2}(|{S}|^{2}+|{P_{2}}|^{2})
  + E_{V}^{2}(|{S+D}|^{2} + |{P_{1}}|^{2})},
\label{kin3}
\vspace{2mm}
\end{eqnarray}
and
\begin{eqnarray}
 S &=& -A_{1}, \nn\\
 D &=& -\frac{p_{c}^{2}}{E_{V}(E_{\Lamc}+m_{\Lamc})}
   (A_{1} - m_{\Lambda_{b}}A_{2}), \nn\\
 P_{1}&=& -\frac{p_{c}}{E_{V}}(\frac{m_{\Lambda_{b}} + m_{\Lamc}}
   {E_{\Lamc} + m_{\Lamc}}B_{1} + m_{\Lambda_{b}}B_{2}), \nn\\
 P_{2} &=& \frac{p_{c}}{E_{\Lamc}+m_{\Lamc}}B_{1},
\label{kin4}
\vspace{2mm}
\end{eqnarray}
where
\begin{eqnarray}
A_1&=&-\frac{G_F}{\sqrt{2}}V_{cb}V^{*}_{UD}a_1 f_V m_V[g_1(m_{V}^{2})
+g_2(m_{V}^{2})(m_{\Lamb}-m_{\Lamc})],  \nn\\
A_2&=&-2\frac{G_F}{\sqrt{2}}V_{cb}V^{*}_{UD}a_1 f_V m_V
g_2(m_{V}^{2}), \nn\\
B_1&=&\frac{G_F}{\sqrt{2}}V_{cb}V^{*}_{UD}a_1 f_V m_V[f_1(m_{V}^{2})
-f_2(m_{V}^{2})(m_{\Lamb}+m_{\Lamc})],  \nn\\
B_2&=&2\frac{G_F}{\sqrt{2}}V_{cb}V^{*}_{UD}a_1 f_V m_V
f_2(m_{V}^{2}).
\label{kin5}
\vspace{2mm}
\end{eqnarray}

\vspace{0.2in}
{\large\bf III. The results}
\vspace{0.2in}

In the B-S equation approach \cite{bs} $\Lambda_Q$ is regarded as the 
bound state of a heavy quark and a light scalar diquark. 
The B-S wave function $\chi_P(p)$ satisfies the following B-S equation
\begin{equation}
\chi_P(p)=S_F(\lambda_1 P+p)\int \frac{\rd^4q}{(2\pi)^4}G(P,p,q)\chi_P(q)
S_D(-\lambda_2 P+p),
\label{9}
\vspace{2mm}
\end{equation}
where $P=m_{\Lambda_{Q}}v$ is the momentum of $\Lambda_Q$,
the two parameters $\lambda_1$ and $\lambda_2$ are defined in terms of the
heavy quark mass $m_Q$ and the light diquark mass $m_D$ in the baryon, 
$\lambda_1=\frac{m_Q}{m_Q+m_D}, \lambda_2=\frac{m_D}{m_Q+m_D}$, $p$ is the 
relative momentum of the two constituents. The momentum of the heavy quark is
$p_1=\lambda_1 P+p$ and that of the diquark is $p_2=-\lambda_2 P+p$.
In eq. (\ref{9}) $S_F$ and $S_D$ are the propagators of the heavy quark and 
the diquark respectively and  $G(P,p,q)$ is the kernel which is the sum of 
the two particle irreducible diagrams. The kernel is assumed  to have
the  form
\begin{equation}
-iG=I\otimes I V_1 +v_{\mu} \otimes (p_2+p'_2)^{\mu} V_2, 
\label{10}
\vspace{2mm}
\end{equation}
where the first term arises from scalar confinement and the second one 
is from one gluon exchange diagram, $p_2$ and $p'_2$ are the momenta of
the diquark. 
For convenience we introduce the longitudinal and transverse momentum variables 
$p_l=v\cdot p -\lambda_2 m_{\Lambda_Q}, p_t=p-(v\cdot p)v$.  
In the heavy quark 
limit the heavy quark is almost on-shell, hence we can make the 
convariant instantaneous approximation $p_l=q_l$ in the kernel.
Thus $V_1$ and $V_2$ are of the following forms
\begin{eqnarray}
V_1&=&\frac{8\pi\kappa}{[(p_t-q_t)^2+\mu^2]^2}-(2\pi)^3
\delta^3  (p_t-q_t)
	\int \frac{\rd^3 k}{(2\pi)^3}\frac{8\pi\kappa}{(k^2+\mu^2)^2}, \nn \\
V_2&=&-\frac{16\pi}{3}
	\frac{\alpha_{s
eff}^{2}Q_{0}^{2}}{[(p_t-q_t)^2+\mu^2][(p_t-q_t)^2+Q_{0}^{2}]}.
\label{11}
\vspace{2mm}
\end{eqnarray}
where the parameter $\kappa$ which describes the strength of linear 
confinement varies from 0.02GeV$^3$ to 0.1GeV$^3$ and
$\alpha_{s eff}$ changes correspondingly. The parameter $Q_{0}^{2}$ 
is introduced since the diquark is not point-like and we use the value
$Q_{0}^{2}=3.2$GeV$^2$ \cite{kroll}.

In the limit $m_Q \rightarrow \infty$ $\chi_P(p)$ is only related
to a scalar function $\phi_P(p)$ which controls the dynamics
\begin{equation}
\chi_P(p)=\phi_P(p)u_{\Lambda_Q}(v,s).
\label{12}
\vspace{2mm}
\end{equation}

Defining $\tilde{\phi}_P(p_t)=\int \frac{\rd p_l}{2\pi} \phi_P(p)$ we have
the B-S equation for $\tilde{\phi}_P(p_t)$
\begin{equation}
\tilde{\phi}_P(p_t)=-\frac{1}{2(E_0-W_p+m_D)W_{p}} 
\int \frac{\rd^3q_t}{(2\pi)^3}(V_1 -2W_p V_2)\tilde{\phi}_P(q_t).
\label{13}
\vspace{2mm}
\end{equation}
where $W_{p}=\sqrt{p_{t}^{2}+m_{D}^{2}}$ and $E_0$ is the binding energy.

The solutions for the B-S wave function $\chi_P(p)$ can be applied 
to obtain the numerical result for the Isgur-Wise function for $\Lambda_b
\rightarrow \Lambda_c$
\begin{equation}
\xi(\omega)=\int
\frac{\rd^4p}{(2\pi)^4}\phi_{P'}(p')\phi_{P}(p)S_{D}^{-1}(p_2).
\label{14}
\vspace{2mm}
\end{equation}
The numerical values for the Isgur-Wise function depend on the parameters
$\kappa$ and $m_D$. We let $m_D$ vary from 650MeV to 800MeV.

In the limit $m_{b,c} \rightarrow \infty$ A and B in eq.(\ref{kin2}) 
are given by the value of the Isgur-Wise function at $m_{P}^{2}$, 
$\xi (m_{P}^{2})$. Then from eq.(\ref{kin1}) we obtain the decay widths,
branching ratios and the asymmetry parameters. The results for different 
final pseudoscalar or vector mesons are listed in Table 1.

\begin{center}
{\bf Table  1.  Predictions for $\Lamb \rightarrow \Lamc P(V)$ in B-S 
approach ($m_{b,c} \rightarrow \infty$)}
\end{center}
\begin{center}
\begin{tabular}{|c|c|c|c|c|c|} 
\hline
  &$\Gamma (10^{10}s^{-1}$) & B ($\%$) &$\alpha$ \\
\hline
$\Lamb^0 \rightarrow \Lamc^+ \pi^-$  &$0.16a_{1}^{2}\sim 0.28a_{1}^{2}$  
&$0.18a_{1}^{2}\sim 0.32a_{1}^{2}$ &-1.000 \\
\hline
$\Lamb^0 \rightarrow \Lamc^+ \rho^-$  &$0.23a_{1}^{2}\sim 0.40a_{1}^{2}$  
&$0.26a_{1}^{2}\sim 0.46a_{1}^{2}$ &-0.899 \\
\hline
$\Lamb^0 \rightarrow \Lamc^+ D_{s}^{-}$  &$0.62a_{1}^{2}\sim 0.93a_{1}^{2}$  
&$0.71a_{1}^{2}\sim 1.06a_{1}^{2}$ &-0.984 \\
\hline
$\Lamb^0 \rightarrow \Lamc^+ D_{s}^{*-}$  &$0.48a_{1}^{2}\sim 0.70a_{1}^{2}$  
&$0.55a_{1}^{2}\sim 0.80a_{1}^{2}$ &-0.423 \\
\hline
$\Lamb^0 \rightarrow \Lamc^+ K^-$  &$0.011a_{1}^{2}\sim 0.020a_{1}^{2}$  
&$0.013a_{1}^{2}\sim 0.023a_{1}^{2}$ &-1.000 \\
\hline
$\Lamb^0 \rightarrow \Lamc^+ K^{*-}$  &$0.012a_{1}^{2}\sim 0.018a_{1}^{2}$  
&$0.014a_{1}^{2}\sim 0.021a_{1}^{2}$ &-0.866 \\
\hline
$\Lamb^0 \rightarrow \Lamc^+ D^{-}$  &$0.022a_{1}^{2}\sim 0.033a_{1}^{2}$  
&$0.025a_{1}^{2}\sim 0.038a_{1}^{2}$ &-0.988 \\
\hline
$\Lamb^0 \rightarrow \Lamc^+ D^{*-}$  &$0.016a_{1}^{2}\sim 0.024a_{1}^{2}$  
&$0.019a_{1}^{2}\sim 0.028a_{1}^{2}$ &-0.463 \\
\hline
\end{tabular}
\end{center}
\vspace{2mm}

It is noted that in Table 1 the range of the prediction values corresponds to
$\kappa$ from 0.02GeV$^3$ to 0.1GeV$^3$. $m_D$ is chosen to 700MeV. The
numerical results for $m_D$ from 650MeV to 800MeV change very little.
Furthermore, in the calculations we use the following decay constants
$$f_\pi=132MeV, \;\; f_K=156MeV, \;\; f_D=200MeV\cite{dai}\cite{sa}, \;\; 
f_{D_s}=241MeV\cite{sa},$$
$$f_\rho=216MeV, \;\; f_{K^*}=f_\rho, \;\; f_D=f_{D^*}, \;\; 
f_{D_s}=f_{D_{s}^{*}}.$$
The masses for $\Lamb^0$ and $\Lamc^+$ are 5.641GeV and 2.285GeV respectively
\cite{data} and the lifetime of $\Lamb$ is $1.14\times 10^{-12}s$\cite{data}.

In ref. \cite{gk} Guo and Kroll use the Drell-Yan type overlap integrals 
for the model hadronic wave functions of $\Lambda_b$ and $\Lambda_c$ to obtain 
the form factors $F_i, G_i (i=1,2,3)$ in eq. (\ref{7}). It is noted from 
the HQET that to the order $1/m_b, 1/m_c$ expansion all the six form factors 
are realted with each other
\begin{eqnarray}
F_1&=&G_1\left[ 1+\left(\frac{1}{m_b}+\frac{1}{m_c}\right)\frac{\bar{\Lambda}}
{1+\omega}\right], \nn\\
F_2&=&G_2=-G_1\frac{1}{m_c}\frac{\bar{\Lambda}}{1+\omega}, \nn\\
F_3&=&-G_3=-G_1\frac{1}{m_b}\frac{\bar{\Lambda}}{1+\omega},
\label{15}
\vspace{2mm}
\end{eqnarray}
where $\bar{\Lambda}$ is the unknown parameter which is the difference between
the mass of the heavy baryon and that of the heavy quark.
Therefore, if one determines one form factor all the six form factors are
known. 

Just as in the case of the B-S approach, $\Lambda_Q$ is regarded as composed
of a heavy quark and a scalar light diquark in IMF.
\begin{equation}
|\Lambda_{Q}(\vec{P},\lambda> = \sqrt{\frac{m_{Q}}{2m_{\Lambda_{Q}}}}
   \int\frac{d^{3}k}{\sqrt{E_{Q}E_{D}}}\Psi_{\Lambda_{Q}}(\vec{k})|
   Q(\vec{P}-\vec{k}),\lambda;D(\vec{k})>,
\label{16}
\vspace{2mm}
\end{equation}	
where color indices have been omitted, $E_{Q}$ and $E_{D}$ are 
the IMF energies of the
heavy quark and scalar-particle, respectively, $\lambda$ represents 
the helicity of the baryon. The renormalization of the wave function is
\begin{equation}
\int{dx_{1}}d^{2}k_{\perp}|\Psi_{\Lambda_{Q}}(x_{1},\vec{k}_{\perp})|^{2} = 1, 
\label{17}
\vspace{2mm}
\end{equation}
where the the longitudinal momentum 
fraction $x_{1}$ carried by the heavy quark and
the heavy quark's transverse momentum corresponding to its parent baryon
$\vec{k}_{\perp}$ are introduced. Obviously, the scalar diquark 
carries $x_{2}=1-x_{1}$ and -$\vec{k}_{\perp}$. The baryon wave function 
$\Psi_{\Lambda_{Q}}(x_{1},\vec{k}_{\perp})$
is a generalization of the Bauer-Stech-Wirbel \cite{bsw} meson
wave function to the quark-diquark case

\begin{equation}
\Psi_{\Lambda_{Q}}(x_{1},\vec{k}_{\perp}) = N_{\Lambda_{Q}}x_{1}x_{2}^3
exp[-b^{2}(\vec{k}_{\perp}^2 + m_{\Lambda_{Q}}^{2}(x_{1}-x_{0})^{2})],
\label{18}
\vspace{2mm}
\end{equation}
where $N_{\Lambda_{Q}}$ is the normalization constant.
The peak position of the wave function 
is at $x_{0}=1-\bar{\Lambda}/m_{\Lambda_{Q}}$ and the width is of the order
$\bar{\Lambda}/m_{\Lambda_{Q}}$. In the calculations 
$\bar{\Lambda}=600$MeV is used. Another parameter b in the wave function
is related to the mean $k_{\perp}$ or the radius of 
the baryon and its precise value
is not known. However, we expect the radius of a heavy baryon
to be smaller than that of proton.
In the following calculations, as in \cite{gk}, we use b=1.77GeV and b=1.18
GeV, corresponding to $<k_{\perp}^{2}>$$^{\frac{1}{2}}$ = 400 MeV and 
$<k_{\perp}^{2}>$$^{\frac{1}{2}}$ = 600MeV respectively.

The form factor, say $F_1$, is related to the overlap integral
of the hadronic wave functions of $\Lambda_b$ and $\Lambda_c$. To the
leading order in $1/m_Q$ expansion it is just the Isgur-Wise function
which has the following form
\begin{equation}
\xi(\omega)= \left(\frac{2}{\omega+1}\right)exp\left(-2{\bar{\Lambda}}^2
b^2\frac{\omega-1}{\omega+1}\right) \frac{K_6(2\bar{\Lambda}
b/\sqrt{\omega+1})}{K_6(\sqrt{2}\bar{\Lambda} b)}.
\label{19}
\vspace{2mm}
\end{equation}
where $K_l$ is defined as 
$$K_l (x)=\int_{-x}^{\infty}\rd z e^{-z^2}(z+x)^l.$$

With the above form of the Isgur-Wise function we obtain the decay widths,
branching ratios and the asymmetry parameters in the hadronic wave function
model. The results  are listed in Table 2(a).

{\bf Table 2(a).  Predictions for $\Lamb \rightarrow \Lamc P(V)$ in hadronic
wave function model ($m_{b,c}\rightarrow \infty$)}
\begin{center}
\begin{tabular}{|c|c|c|c|c|c|} 
\hline
  &$\Gamma (10^{10}s^{-1}$) & B ($\%$) &$\alpha$ \\
\hline
$\Lamb^0 \rightarrow \Lamc^+ \pi^-$  &$0.064a_{1}^{2}\sim 0.12a_{1}^{2}$  
&$0.073a_{1}^{2}\sim 0.14a_{1}^{2}$ &-1.000 \\
\hline
$\Lamb^0 \rightarrow \Lamc^+ \rho^-$  &$0.098a_{1}^{2}\sim 0.18a_{1}^{2}$  
&$0.11a_{1}^{2}\sim 0.20a_{1}^{2}$ &-0.899 \\
\hline
$\Lamb^0 \rightarrow \Lamc^+ D_{s}^{-}$  &$0.33a_{1}^{2}\sim 0.50a_{1}^{2}$  
&$0.38a_{1}^{2}\sim 0.57a_{1}^{2}$ &-0.984 \\
\hline
$\Lamb^0 \rightarrow \Lamc^+ D_{s}^{*-}$  &$0.27a_{1}^{2}\sim 0.40a_{1}^{2}$  
&$0.31a_{1}^{2}\sim 0.45a_{1}^{2}$ &-0.423 \\
\hline
$\Lamb^0 \rightarrow \Lamc^+ K^-$  &$0.0047a_{1}^{2}\sim 0.0086a_{1}^{2}$  
&$0.0054a_{1}^{2}\sim 0.0098a_{1}^{2}$ &-1.000 \\
\hline
$\Lamb^0 \rightarrow \Lamc^+ K^{*-}$  &$0.0053a_{1}^{2}\sim 0.0093a_{1}^{2}$  
&$0.0060a_{1}^{2}\sim 0.011a_{1}^{2}$ &-0.866 \\
\hline
$\Lamb^0 \rightarrow \Lamc^+ D^{-}$  &$0.011a_{1}^{2}\sim 0.017a_{1}^{2}$  
&$0.013a_{1}^{2}\sim 0.020a_{1}^{2}$ &-0.988 \\
\hline
$\Lamb^0 \rightarrow \Lamc^+ D^{*-}$  &$0.0088a_{1}^{2}\sim 0.013a_{1}^{2}$  
&$0.010a_{1}^{2}\sim 0.015a_{1}^{2}$ &-0.463 \\
\hline
\end{tabular}
\end{center}
\vspace{2mm}

By using eq. (\ref{15}) all the six form factors are obtained to the order
$1/m_b, 1/m_c$
\begin{eqnarray}
F_1&=&\xi(\omega)\left[ 1+\bar{\Lambda}\left(\frac{1}{m_b}+\frac{1}{m_c}
\right)\left[g(\omega)+\frac{1}{1+\omega}\right]\right], \nn\\
G_1&=&\xi(\omega)\left[1+\bar{\Lambda}\left(\frac{1}{m_b}+\frac{1}{m_c}
\right)g(\omega)\right], \nn\\
F_2&=&G_2=\frac{m_b}{m_c}F_3=-\frac{m_b}{m_c}G_3=-\frac{\bar{\Lambda}}{m_c}
\frac{\xi(\omega)}{1+\omega},
\label{20}
\vspace{2mm}
\end{eqnarray}
where $g(\omega)$ is the $1/m_Q$ correction term
\begin{equation}
g(\omega)=\frac{1}{\sqrt{2}\bar{\Lambda}b}\left[\frac{K_7(\sqrt{2}\bar{\Lambda}
b)}{K_6(\sqrt{2}\bar{\Lambda} b)}-\sqrt{\frac{2}{1+\omega}}
\frac{K_7(2\bar{\Lambda}
b/\sqrt{\omega+1})}{K_6(2\bar{\Lambda}b/\sqrt{\omega+1})}\right].
\label{21}
\vspace{2mm}
\end{equation}

After including the $1/m_Q$ corrections we obtain Table 2(b).

{\bf Table 2(b).  Predictions for $\Lamb \rightarrow \Lamc P(V)$ in hadronic 
wave function model (with $1/m_Q$ corrections)}
\begin{center}
\begin{tabular}{|c|c|c|c|c|c|} 
\hline
  &$\Gamma (10^{10}s^{-1}$) & B ($\%$) &$\alpha$ \\
\hline
$\Lamb^0 \rightarrow \Lamc^+ \pi^-$  &$0.077a_{1}^{2} \sim 0.34a_{1}^{2}$  
&$0.088a_{1}^{2}\sim 0.39a_{1}^{2}$ &-1.000 \\
\hline
$\Lamb^0 \rightarrow \Lamc^+ \rho^-$  &$0.11a_{1}^{2}\sim 0.48a_{1}^{2}$  
&$0.13a_{1}^{2}\sim 0.55a_{1}^{2}$ &-0.890 $\sim$ 0.893 \\
\hline
$\Lamb^0 \rightarrow \Lamc^+ D_{s}^{-}$  &$0.37a_{1}^{2}\sim 1.10a_{1}^{2}$  
&$0.42a_{1}^{2}\sim 1.25a_{1}^{2}$ &-0.984 \\
\hline
$\Lamb^0 \rightarrow \Lamc^+ D_{s}^{*-}$  &$0.30a_{1}^{2}\sim 0.84a_{1}^{2}$  
&$0.35a_{1}^{2}\sim 0.96a_{1}^{2}$ &-0.390 $\sim$ 0.403 \\
\hline
$\Lamb^0 \rightarrow \Lamc^+ K^-$  &$0.0059a_{1}^{2}\sim 0.024a_{1}^{2}$  
&$0.0063a_{1}^{2}\sim 0.027a_{1}^{2}$ &-1.000 \\
\hline
$\Lamb^0 \rightarrow \Lamc^+ K^{*-}$  &$0.0061a_{1}^{2}\sim 0.025a_{1}^{2}$  
&$0.0070a_{1}^{2}\sim 0.029a_{1}^{2}$ &-0.856 $\sim$ 0.859 \\
\hline
$\Lamb^0 \rightarrow \Lamc^+ D^{-}$  &$0.013a_{1}^{2}\sim 0.039a_{1}^{2}$  
&$0.014a_{1}^{2}\sim 0.045a_{1}^{2}$ &-0.988 \\
\hline
$\Lamb^0 \rightarrow \Lamc^+ D^{*-}$  &$0.010a_{1}^{2}\sim 0.029a_{1}^{2}$  
&$0.011a_{1}^{2}\sim 0.033a_{1}^{2}$ &-0.431 $\sim$ 0.440 \\
\hline
\end{tabular}
\end{center}
\vspace{2mm}

The Isgur-Wise functions for $\Lamb \rightarrow \Lamc$ are also calculated
in other models. From the soliton model 
Jenkins, Manohar and Wise get the following form\cite{soliton}
\begin{equation}
\xi(\omega)= 0.99 exp[-1.3(\omega-1)].
\label{22}
\vspace{2mm}
\end{equation}

The MIT bag model calculation by Sadzikowski and Zalewski
\cite{mit}) gives the following result
\begin{equation}
\xi(\omega)= \left(\frac{2}{\omega+1}\right)^{3.5+1.2/\omega}.
\label{23}
\vspace{2mm}
\end{equation}

In ref. \cite{cheng1} the authors calculated the Cabibbo-favored two-body
nonleptonic decays in the nonrelativistic quark model. The advantage
of this approach is that the daughter baryon can also be light. So this
approach gives predictions for many processes. The weak transition form
factors for $\Lamb$ to the daughter baryon are first calculated at the
zero recoil point at which the daughter baryon is also at rest in the
rest frame of $\Lamb$. Then the $q^2$ dependence is introduced by the
assumption of the dipole behavior of the form factors. It is noted that
in this approach the $1/m_Q$ corrections are included.

The decay widths and asymmetry parameters from 
these three models are listed in Table 3.

{\bf Table 3.  Predictions for $\Lamb \rightarrow \Lamc P(V)$ in
soliton ($\Gamma_1$, $\alpha_1$)($m_{b,c}\rightarrow \infty$)
and MIT bag model ($\Gamma_2$, $\alpha_2$) 
($m_{b,c}\rightarrow \infty$) and nonrelativistic quark model
($\Gamma_3$, $\alpha_3$) ($1/m_Q$ corrections included)}
\begin{center}
\begin{tabular}{|c|c|c|c|c|c|c|c|} 
\hline
  &$\Gamma_1 (10^{10}s^{-1})$ &$\Gamma_2 (10^{10}s^{-1})$  
&$\Gamma_3 (10^{10}s^{-1})$&$\alpha_1=\alpha_2$ &$\alpha_3$ \\
\hline
$\Lamb^0 \rightarrow \Lamc^+ \pi^-$  &$0.36a_{1}^{2}$  
&$0.20a_{1}^{2}$ &$0.31a_{1}^{2}$ &-1.000  &-0.99\\
\hline
$\Lamb^0 \rightarrow \Lamc^+ \rho^-$  &$0.50a_{1}^{2}$  
&$0.29a_{1}^{2}$ &$0.44a_{1}^{2}$ &-0.899 &-0.88\\
\hline
$\Lamb^0 \rightarrow \Lamc^+ D_{s}^{-}$  &$1.10a_{1}^{2}$  
&$0.72a_{1}^{2}$ &$0.93a_{1}^{2}$ &-0.984 &-0.99\\
\hline
$\Lamb^0 \rightarrow \Lamc^+ D_{s}^{*-}$  &$0.82a_{1}^{2}$  
&$0.55a_{1}^{2}$ &$0.74a_{1}^{2}$ &-0.423 &-0.36\\
\hline
$\Lamb^0 \rightarrow \Lamc^+ K^-$  &$0.025a_{1}^{2}$  
&$0.015a_{1}^{2}$ & &-1.000 &\\
\hline
$\Lamb^0 \rightarrow \Lamc^+ K^{*-}$  &$0.026a_{1}^{2}$  
&$0.016a_{1}^{2}$ & &-0.866 & \\
\hline
$\Lamb^0 \rightarrow \Lamc^+ D^{-}$  &$0.039a_{1}^{2}$  
&$0.025a_{1}^{2}$ & &-0.988 & \\
\hline
$\Lamb^0 \rightarrow \Lamc^+ D^{*-}$  &$0.029a_{1}^{2}$  
&$0.019a_{1}^{2}$ & &-0.463 & \\
\hline
\end{tabular}
\end{center}
\vspace{2mm}

\vspace{0.2in}
{\large\bf IV. Summary and discussions}
\vspace{0.2in}

	From the numerical results in Section III we can see the following
conclusions. 

	(i) In the heavy quark limit $m_{b,c} \rightarrow \infty$ the decay
widths are determined by the Isgur-Wise function. The asymmetry parameter does
not depend the form of the Isgur-Wise function since it is canceled in
eqs. (\ref{kin1}) and (\ref{kin3}). The results for $m_{b,c} \rightarrow 
\infty$ in different models are listed in Table 1, Table 2(a) and 
$\Gamma_1, \Gamma_2, \alpha_1 (=\alpha_2)$ in Table 3. We can see that 
only the B-S approach is consistent with the MIT bag model, but they are
not consistent with the hadronic wave function model and the soliton model.
This is because the Isgur-Wise functions in these models are different.
It is noted that the asymmetry parameter is independent of the models
we used in the leading order of $1/m_Q$ expansion. The reason is that 
to this order we only have the Isgur-Wise function and it is canceled 
in the expression of the asymmetry parameter.

	(ii) In the hadronic wave function model and the nonrelativistic
quark model the $1/m_Q$ corrections are taken into account. We can see from
Table 2(b) and $\Gamma_3$ in Table 3 that the decay widths in these two 
models are consistent with each other. Furthermore the asymmetry parameter
$\alpha$ depends on the form factors to the order $1/m_Q$.

	(iii) In our B-S and hadronic wave function models the predictions
vary in a large range because of the uncertainty parameter $\kappa$ in
the B-S approach and $<k_{\perp}>$ in the hadronic wave function model. 
The forthcoming experimental data are needed to determine these parameters.

	(iv) From the hadronic wave function model we can see that the $1/m_Q$
corrections could be important. This is because the the correction from the
$1/m_c$ is of the order $\bar{\Lambda}/m_c$. Taking $\bar{\Lambda} \sim 600$MeV
and $m_c \sim 1.5$GeV the $1/m_c$ correction could be about 50$\%$. In the
hadronic wave function model the $1/m_Q$ correction makes the decay widths
bigger. It is not clear yet in other models whether the $1/m_Q$ correction will
make the results bigger or smaller. This needs further study.

	(v) Our calculations are carried out in the factorization approach.
In the bottom baryon case it is a good approximation. However, how large the
nonfactorization contributions are is still a problem. In ref. \cite{cheng0}
it is estimated that the W-exchange diagram contribution is about 15$\%$
of the spectator diagram. In a recent paper \cite{korner} the nonfactorization
contributions are calculated in a relativistic quark model and they
amount up to about 30$\%$ of the factorization contribution
in $b \rightarrow c$ transition. 
In our approach we do not calculate the nonfactorization contributions. 
However, in the following, we try to 
give a discussion about the consistency of the 
factorization approach with the relation between 
$\Lambda_b \rightarrow \Lambda_c D_s$ and $\Lambda_b \rightarrow \Lambda_c
D_{s}^{*}$ given by the heavy quark symmetry.
	
Because of the heavy quark symmetry in the limit $m_{b,c} \rightarrow \infty$
the amplitudes for $\Lambda_b \rightarrow \Lambda_c D_s$ and $\Lambda_b 
\rightarrow \Lambda_c D_{s}^{*}$ are \cite{grinstein}

\begin{equation}
{\cal A}(\Lambda_b (v_{\Lambda_b})\rightarrow \Lambda_c (v_{\Lambda_c})
D_s(\bar{v}))=
i\bar{u}(v_{\Lambda_c})(S+P\gamma_5)u(v_{\Lambda_b}),
\label{hqs1}
\vspace{2mm}
\end{equation}
\begin{equation}
{\cal A}(\Lambda_b (v_{\Lambda_b})\rightarrow \Lambda_c (v_{\Lambda_c})
D_{s}^{*}(\bar{v}))=
2\bar{u}(v_{\Lambda_c})(1+\gamma_5)[(A+2Bv_{\Lambda_b}\cdot\bar{v})
{\rlap / \epsilon^{*}}-2B(\epsilon^{*}\cdot v_{\Lambda_b})\rlap /{\bar{v}}+B
\rlap /{\bar{v}} 
{\rlap /\epsilon^{*}}]u(v_{\Lambda_b}),
\label{hqs2}
\vspace{2mm}
\end{equation}
where $\bar{v}$ is the velocity of $D_s$ or $D_{s}^{*}$, and
\begin{eqnarray}
A&=&\frac{S-P}{4}, \nn\\
B&=&-\frac{m_b}{m_c}\frac{S-P}{4}-\frac{S+P}{4}.
\label{hqs3}
\vspace{2mm}
\end{eqnarray}
The equations (\ref{hqs1}) and (\ref{hqs2}) are general forms in the limit 
$m_{b,c} \rightarrow \infty$ and hence no factorization assumption is made.

To show the consistency of the factorization approach with the eqs.
(\ref{hqs1}, \ref{hqs2}) we 
first use eq. (\ref{hqs1}) to find the expressions for
S and P in the factorization approximation. Comparing the amplitudes for
$\Lambda_b \rightarrow \Lambda_c D_{s}$ in 
eqs. (\ref{4}) and
(\ref{hqs1}) in the leading order of $1/m_Q$ expansion we find
\begin{eqnarray}
S&=&\frac{G_F}{\sqrt{2}}V_{cb}V^{*}_{UD}a_1f_{D_s}
(m_{\Lambda_b}-m_{\Lambda_c})\xi(\omega), \nn \\
P&=&\frac{G_F}{\sqrt{2}}V_{cb}V^{*}_{UD}a_1f_{D_s}
(m_{\Lambda_b}+m_{\Lambda_c})\xi(\omega).
\label{hqs4}
\vspace{2mm}
\end{eqnarray}
Hence from eq. (\ref{hqs3}) we find that in the factorization approximation
\begin{eqnarray}
A&=&-\frac{1}{2}\frac{G_F}{\sqrt{2}}V_{cb}V^{*}_{UD}a_1f_{D_s}
m_{\Lambda_c}\xi(\omega), \nn \\
B&=&0.
\label{hqs5}
\vspace{2mm}
\end{eqnarray}
Substituting eq. (\ref{hqs5}) into eq. (\ref{hqs2}) we get the decay width
$\tilde{\Gamma}(\Lambda_b \rightarrow \Lambda_c D_{s}^{*})$. On the other
hand, if we simply calculate the decay width for $\Lambda_b \rightarrow 
\Lambda_c D_{s}^{*}$ from eq. (\ref{4}) in the factorization approach we
get $\Gamma^{fac}(\Lambda_b \rightarrow \Lambda_c D_{s}^{*})$. If the
factorization approach is completely consistent with the relations
between the decay widths of $\Lambda_b \rightarrow \Lambda_c D_{s}$
and $\Lambda_b \rightarrow \Lambda_c D_{s}^{*}$ provided by the heavy quark
symmetry then 
$\Gamma^{fac}(\Lambda_b \rightarrow \Lambda_c D_{s}^{*})$ should be equal to
$\tilde{\Gamma}(\Lambda_b \rightarrow \Lambda_c D_{s}^{*})$. We find that
the ratio of $\Gamma^{fac}(\Lambda_b \rightarrow \Lambda_c D_{s}^{*})$
and $\tilde{\Gamma}(\Lambda_b \rightarrow \Lambda_c D_{s}^{*})$ is 0.86
which is close to 1.
This ratio is independent of the nonperturbative QCD models  
since the Isgur-Wise function
is canceled. So it seems that the factorization approach is satisfactorily
consistent with the general requirements from the heavy quark symmetry.

\vspace{1cm}

\noindent {\bf Acknowledgment}:
\vspace{0.2in}

This work was supported in part by the Australian Research Council and
the National Science Foundation of China.

\end{document}